
\input harvmac
\input tables

\overfullrule=0pt
 

\def\B{{\scriptscriptstyle B}}
\def\C{{\scriptscriptstyle C}}

\def\I{{\scriptscriptstyle I}}
\def\J{{\scriptscriptstyle J}}

\def\L{{\scriptscriptstyle L}}

\def\Q{{\scriptscriptstyle Q}}
\def\R{{\scriptscriptstyle R}}
\def\S{{\scriptscriptstyle S}}

\def\V{{\scriptscriptstyle V}}

\def\Y{{\scriptscriptstyle Y}}


\def\Idot{{\scriptscriptstyle \dot{I}}}
\def\Jdot{{\scriptscriptstyle \dot{J}}}


\def\CN{{\cal N}}

 
\def\a{\alpha}
\def\b{\beta}
\def\c{\gamma}

\def\e{\epsilon}


\def\half{{1 \over 2}}
\def\quarter{{1 \over 4}}

\def\third{{1 \over 3}}


\def\cdot{{\dot{\gamma}}}

\def\bar#1{\overline{#1}}

\def\ccdot{\hbox{\kern-.1em$\cdot$\kern-.1em}}
\def\dash{{\> \over \>}} 		

\def\dabc{d_{\a\b\c}}
\def\dbar{\bar{d}}
\def\Dbar{\bar{D}}

\def\Gdual{{\widetilde G}}
\def\gtap{\raise.3ex\hbox{$>$\kern-.75em\lower1ex\hbox{$\sim$}}}

\def\Hdual{{\widetilde H}}

\def\Idual{{\widetilde I}}

\def\L{{\scriptscriptstyle L}}

\def\LS{L_\S}
\def\LC{L_\C}
\def\ltap{\raise.3ex\hbox{$<$\kern-.75em\lower1ex\hbox{$\sim$}}}
\def\Nc{N_c}
\def\Ncdual{{\tilde N_c}}
\def\Ncpdual{{{{{\tilde N_c}'}}}}
\def\Nf{{N_f}}
\def\OCV{O_{\C\V}}
\def\OSC{O_{\S\C}}
\def\OSV{O_{\S\V}}
\def\Pf{{\rm Pf}\>}
\def\R{{\scriptscriptstyle R}}
\def\Qp{{Q'}}
\def\qbar{{\bar{q}}}

\def\qp{{q'}}
\def\qpp{{q''}}
\def\qppp{{q'''}}

\def\sp{\>}

\def\therefore{{\hbox{..}\kern-.43em \raise.5ex \hbox{.}}\>\>}

\def\Tr{{\rm Tr}}
\def\ubar{\bar{u}}
\def\Vslash{V\hskip-0.75 em / \hskip+0.30 em}
\def\Wdual{{\widetilde W}}
\def\Wdualtree{{\widetilde W}_{\rm tree}}
\def\Wtree{W_{\rm tree}}

\newdimen\pmboffset
\pmboffset 0.022em
\def\oldpmb#1{\setbox0=\hbox{#1}%
 \copy0\kern-\wd0 \kern\pmboffset\raise 1.732\pmboffset\copy0\kern-\wd0
 \kern\pmboffset\box0}
\def\pmb#1{\mathchoice{\oldpmb{$\displaystyle#1$}}{\oldpmb{$\textstyle#1$}}
      {\oldpmb{$\scriptstyle#1$}}{\oldpmb{$\scriptscriptstyle#1$}}}


\def\fund{  {\vcenter  {\vbox  
              {\hrule height.6pt
               \hbox {\vrule width.6pt  height5pt  
                      \kern5pt 
                      \vrule width.6pt  height5pt }
               \hrule height.6pt}
                         }
                   }
           }

\def\antifund{  \overline{ {\vcenter  {\vbox
              {\hrule height.6pt
               \hbox {\vrule width.6pt  height5pt
                      \kern5pt
                      \vrule width.6pt  height5pt }
               \hrule height.6pt}
                         }
                   } }
           }
 
\def\sym{  {\vcenter  {\vbox  
              {\hrule height.6pt
               \hbox {\vrule width.6pt  height5pt  
                      \kern5pt 
                      \vrule width.6pt  height5pt 
                      \kern5pt
                      \vrule width.6pt height5pt}
               \hrule height.6pt}
                         }
              }
           }

\def\symbar{  \overline{ {\vcenter  {\vbox
              {\hrule height.6pt
               \hbox {\vrule width.6pt  height5pt
                      \kern5pt
                      \vrule width.6pt  height5pt
                      \kern5pt
                      \vrule width.6pt height5pt}
               \hrule height.6pt}
                         }
              }
           }  }

\def\symthree{  {\vcenter  {\vbox  
              {\hrule height.6pt
               \hbox {\vrule width.6pt  height5pt  
                      \kern5pt 
                      \vrule width.6pt  height5pt 
                      \kern5pt
                      \vrule width.6pt  height5pt 
                      \kern5pt
                      \vrule width.6pt height5pt}
               \hrule height.6pt}
                         }
              }
           }
\def\symfour{  {\vcenter  {\vbox  
              {\hrule height.6pt
               \hbox {\vrule width.6pt  height5pt  
                      \kern5pt 
                      \vrule width.6pt  height5pt 
                      \kern5pt
                      \vrule width.6pt  height5pt 
                      \kern5pt
                      \vrule width.6pt  height5pt 
                      \kern5pt
                      \vrule width.6pt height5pt}
               \hrule height.6pt}
                         }
              }
           }

\def\anti{ {\vcenter  {\vbox  
              {\hrule height.6pt
               \hbox {\vrule width.6pt  height5pt  
                      \kern5pt 
                      \vrule width.6pt  height5pt }
               \hrule height.6pt
               \hbox {\vrule width.6pt  height5pt  
                      \kern5pt 
                      \vrule width.6pt  height5pt }
               \hrule height.6pt}
                         }
              }
           }
\def\twoone{ 
              {\vcenter  {\vbox  
              {\hrule height.6pt
               \hbox {\vrule width.6pt  height5pt  
                      \kern5pt 
                      \vrule width.6pt  height5pt 
                      \kern5pt
                      \vrule width.6pt height5pt}
               \hrule height.6pt
               \hbox {\vrule width.6pt  height5pt  
                      \kern5pt 
                      \vrule width.6pt  height5pt }
               \hrule height.6pt width3.8pt}
                         }
              }
           }
\def\twotwo{ 
              {\vcenter  {\vbox  
              {\hrule height.6pt
               \hbox {\vrule width.6pt  height5pt  
                      \kern5pt 
                      \vrule width.6pt  height5pt 
                      \kern5pt
                      \vrule width.6pt height5pt}
               \hrule height.6pt
               \hbox {\vrule width.6pt  height5pt  
                      \kern5pt 
                      \vrule width.6pt  height5pt 
                      \kern5pt
                      \vrule width.6pt height5pt}
               \hrule height.6pt}
                         }
              }
           }

\def\threethree{ 
              {\vcenter  {\vbox  
              {\hrule height.6pt
               \hbox {\vrule width.6pt  height5pt  
                      \kern5pt 
                      \vrule width.6pt  height5pt 
                      \kern5pt 
                      \vrule width.6pt  height5pt 
                      \kern5pt
                      \vrule width.6pt height5pt}
               \hrule height.6pt
               \hbox {\vrule width.6pt  height5pt  
                      \kern5pt 
                      \vrule width.6pt  height5pt 
                      \kern5pt
                      \vrule width.6pt  height5pt 
                      \kern5pt
                      \vrule width.6pt height5pt}
               \hrule height.6pt }
                         }
              }
           }

\def\fourfourfour{ 
              {\vcenter  {\vbox  
              {\hrule height.6pt
               \hbox {\vrule width.6pt  height5pt  
                      \kern5pt 
                      \vrule width.6pt  height5pt 
                      \kern5pt 
                      \vrule width.6pt  height5pt 
                      \kern5pt 
                      \vrule width.6pt  height5pt 
                      \kern5pt
                      \vrule width.6pt height5pt}
               \hrule height.6pt
               \hbox {\vrule width.6pt  height5pt  
                      \kern5pt 
                      \vrule width.6pt  height5pt 
                      \kern5pt 
                      \vrule width.6pt  height5pt 
                      \kern5pt 
                      \vrule width.6pt  height5pt 
                      \kern5pt
                      \vrule width.6pt height5pt}
               \hrule height.6pt
               \hbox {\vrule width.6pt  height5pt  
                      \kern5pt 
                      \vrule width.6pt  height5pt 
                      \kern5pt 
                      \vrule width.6pt  height5pt 
                      \kern5pt 
                      \vrule width.6pt  height5pt 
                      \kern5pt
                      \vrule width.6pt height5pt}
               \hrule height.6pt}
              }
           }}

\def\twotwotwo{ 
              {\vcenter  {\vbox  
              {\hrule height.6pt
               \hbox {\vrule width.6pt  height5pt  
                      \kern5pt 
                      \vrule width.6pt  height5pt 
                      \kern5pt 
                      \vrule width.6pt  height5pt }
               \hrule height.6pt
               \hbox {\vrule width.6pt  height5pt  
                      \kern5pt 
                      \vrule width.6pt  height5pt 
                      \kern5pt 
                      \vrule width.6pt  height5pt }
               \hrule height.6pt
               \hbox {\vrule width.6pt  height5pt  
                      \kern5pt 
                      \vrule width.6pt  height5pt 
                      \kern5pt 
                      \vrule width.6pt  height5pt }
               \hrule height.6pt }
                         }
              }
           }

\def\twotwotwobar{  \overline
              {\vcenter  {\vbox  
              {\hrule height.6pt
               \hbox {\vrule width.6pt  height5pt  
                      \kern5pt 
                      \vrule width.6pt  height5pt 
                      \kern5pt 
                      \vrule width.6pt  height5pt }
               \hrule height.6pt
               \hbox {\vrule width.6pt  height5pt  
                      \kern5pt 
                      \vrule width.6pt  height5pt 
                      \kern5pt 
                      \vrule width.6pt  height5pt }
               \hrule height.6pt
               \hbox {\vrule width.6pt  height5pt  
                      \kern5pt 
                      \vrule width.6pt  height5pt 
                      \kern5pt 
                      \vrule width.6pt  height5pt }
               \hrule height.6pt }
                         }
              }
            }

\def\threethreethree{ 
              {\vcenter  {\vbox  
              {\hrule height.6pt
               \hbox {\vrule width.6pt  height5pt  
                      \kern5pt 
                      \vrule width.6pt  height5pt 
                      \kern5pt 
                      \vrule width.6pt  height5pt 
                      \kern5pt 
                      \vrule width.6pt  height5pt }
               \hrule height.6pt
               \hbox {\vrule width.6pt  height5pt  
                      \kern5pt 
                      \vrule width.6pt  height5pt 
                      \kern5pt 
                      \vrule width.6pt  height5pt 
                      \kern5pt 
                      \vrule width.6pt  height5pt }
               \hrule height.6pt
               \hbox {\vrule width.6pt  height5pt  
                      \kern5pt 
                      \vrule width.6pt  height5pt 
                      \kern5pt 
                      \vrule width.6pt  height5pt 
                      \kern5pt 
                      \vrule width.6pt  height5pt } }
               \hrule height.6pt }
              }
           }

\def\threethreethreethree{ 
              {\vcenter  {\vbox  
              {\hrule height.6pt
               \hbox {\vrule width.6pt  height5pt  
                      \kern5pt 
                      \vrule width.6pt  height5pt 
                      \kern5pt 
                      \vrule width.6pt  height5pt 
                      \kern5pt 
                      \vrule width.6pt  height5pt }
               \hrule height.6pt
               \hbox {\vrule width.6pt  height5pt  
                      \kern5pt 
                      \vrule width.6pt  height5pt 
                      \kern5pt 
                      \vrule width.6pt  height5pt 
                      \kern5pt 
                      \vrule width.6pt  height5pt }
               \hrule height.6pt
               \hbox {\vrule width.6pt  height5pt  
                      \kern5pt 
                      \vrule width.6pt  height5pt 
                      \kern5pt 
                      \vrule width.6pt  height5pt 
                      \kern5pt 
                      \vrule width.6pt  height5pt }
               \hrule height.6pt
               \hbox {\vrule width.6pt  height5pt  
                      \kern5pt 
                      \vrule width.6pt  height5pt 
                      \kern5pt 
                      \vrule width.6pt  height5pt 
                      \kern5pt 
                      \vrule width.6pt  height5pt } }
               \hrule height.6pt }
              }
           }

\def\threethreethreethreebar{  \overline 
              {\vcenter  {\vbox  
              {\hrule height.6pt
               \hbox {\vrule width.6pt  height5pt  
                      \kern5pt 
                      \vrule width.6pt  height5pt 
                      \kern5pt 
                      \vrule width.6pt  height5pt 
                      \kern5pt 
                      \vrule width.6pt  height5pt }
               \hrule height.6pt
               \hbox {\vrule width.6pt  height5pt  
                      \kern5pt 
                      \vrule width.6pt  height5pt 
                      \kern5pt 
                      \vrule width.6pt  height5pt 
                      \kern5pt 
                      \vrule width.6pt  height5pt }
               \hrule height.6pt
               \hbox {\vrule width.6pt  height5pt  
                      \kern5pt 
                      \vrule width.6pt  height5pt 
                      \kern5pt 
                      \vrule width.6pt  height5pt 
                      \kern5pt 
                      \vrule width.6pt  height5pt }
               \hrule height.6pt
               \hbox {\vrule width.6pt  height5pt  
                      \kern5pt 
                      \vrule width.6pt  height5pt 
                      \kern5pt 
                      \vrule width.6pt  height5pt 
                      \kern5pt 
                      \vrule width.6pt  height5pt } }
               \hrule height.6pt }
              }
            }

\def\threethreethreetwo{ 
              {\vcenter  {\vbox  
              {\hrule height.6pt
               \hbox {\vrule width.6pt  height5pt  
                      \kern5pt 
                      \vrule width.6pt  height5pt 
                      \kern5pt 
                      \vrule width.6pt  height5pt 
                      \kern5pt 
                      \vrule width.6pt  height5pt }
               \hrule height.6pt
               \hbox {\vrule width.6pt  height5pt  
                      \kern5pt 
                      \vrule width.6pt  height5pt 
                      \kern5pt 
                      \vrule width.6pt  height5pt 
                      \kern5pt 
                      \vrule width.6pt  height5pt }
               \hrule height.6pt
               \hbox {\vrule width.6pt  height5pt  
                      \kern5pt 
                      \vrule width.6pt  height5pt 
                      \kern5pt 
                      \vrule width.6pt  height5pt 
                      \kern5pt 
                      \vrule width.6pt  height5pt }
               \hrule height.6pt
               \hbox {\vrule width.6pt  height5pt  
                      \kern5pt 
                      \vrule width.6pt  height5pt 
                      \kern5pt 
                      \vrule width.6pt  height5pt } }
               \hrule height.6pt width11.8pt }
              }
           }

\def\threethreethreeone{ 
              {\vcenter  {\vbox  
              {\hrule height.6pt
               \hbox {\vrule width.6pt  height5pt  
                      \kern5pt 
                      \vrule width.6pt  height5pt 
                      \kern5pt 
                      \vrule width.6pt  height5pt 
                      \kern5pt 
                      \vrule width.6pt  height5pt }
               \hrule height.6pt
               \hbox {\vrule width.6pt  height5pt  
                      \kern5pt 
                      \vrule width.6pt  height5pt 
                      \kern5pt 
                      \vrule width.6pt  height5pt 
                      \kern5pt 
                      \vrule width.6pt  height5pt }
               \hrule height.6pt
               \hbox {\vrule width.6pt  height5pt  
                      \kern5pt 
                      \vrule width.6pt  height5pt 
                      \kern5pt 
                      \vrule width.6pt  height5pt 
                      \kern5pt 
                      \vrule width.6pt  height5pt }
               \hrule height.6pt
               \hbox {\vrule width.6pt  height5pt  
                      \kern5pt 
                      \vrule width.6pt  height5pt } }
               \hrule height.6pt width 6.1pt}
              }
           }

\def\twotwoone{ 
              {\vcenter  {\vbox  
              {\hrule height.6pt
               \hbox {\vrule width.6pt  height5pt  
                      \kern5pt 
                      \vrule width.6pt  height5pt 
                      \kern5pt 
                      \vrule width.6pt  height5pt }
               \hrule height.6pt
               \hbox {\vrule width.6pt  height5pt  
                      \kern5pt 
                      \vrule width.6pt  height5pt 
                      \kern5pt 
                      \vrule width.6pt  height5pt }
               \hrule height.6pt
               \hbox {\vrule width.6pt  height5pt  
                      \kern5pt 
                      \vrule width.6pt  height5pt }
               \hrule height.6pt width6.2pt}
                         }
              }
           }

\def\four{  {\vcenter  {\vbox
              {\hrule height.5pt
               \hbox {\vrule width.5pt  height4pt
                      \kern4pt
                      \vrule width.5pt  height4pt
                      \kern4pt
                      \vrule width.5pt  height4pt
                      \kern4pt
                      \vrule width.5pt  height4pt
                      \kern4pt
                      \vrule width.5pt height4pt}
               \hrule height.5pt}
                         }
              }
           }

\def\fourtwotwo{
              {\vcenter  {\vbox  
              {\hrule height.6pt
               \hbox {\vrule width.6pt  height5pt  
                      \kern5pt 
                      \vrule width.6pt  height5pt 
                      \kern5pt 
                      \vrule width.6pt  height5pt 
                      \kern5pt 
                      \vrule width.6pt  height5pt 
                      \kern5pt
                      \vrule width.6pt height5pt}
               \hrule height.6pt
               \hbox {\vrule width.6pt  height5pt  
                      \kern5pt 
                      \vrule width.6pt  height5pt 
                      \kern5pt
                      \vrule width.6pt height5pt}
               \hrule height.6pt width12.2pt
                 \hbox {\vrule width.6pt  height5pt  
                      \kern5pt 
                      \vrule width.6pt  height5pt 
                      \kern5pt
                      \vrule width.6pt height5pt}
               \hrule height.6pt width12.2pt}                       }
              }
           }

\def\sixsixsix{ 
              {\vcenter  {\vbox  
              {\hrule height.6pt
               \hbox {\vrule width.6pt  height5pt  
                      \kern5pt 
                      \vrule width.6pt  height5pt 
                      \kern5pt 
                      \vrule width.6pt  height5pt 
                      \kern5pt 
                      \vrule width.6pt  height5pt 
                      \kern5pt 
                      \vrule width.6pt  height5pt 
                      \kern5pt 
                      \vrule width.6pt  height5pt 
                      \kern5pt
                      \vrule width.6pt height5pt}
               \hrule height.6pt
               \hbox {\vrule width.6pt  height5pt  
                      \kern5pt 
                      \vrule width.6pt  height5pt 
                      \kern5pt 
                      \vrule width.6pt  height5pt 
                      \kern5pt 
                      \vrule width.6pt  height5pt 
                      \kern5pt 
                      \vrule width.6pt  height5pt 
                      \kern5pt 
                      \vrule width.6pt  height5pt 
                      \kern5pt
                      \vrule width.6pt height5pt}
               \hrule height.6pt
               \hbox {\vrule width.6pt  height5pt  
                      \kern5pt 
                      \vrule width.6pt  height5pt 
                      \kern5pt 
                      \vrule width.6pt  height5pt 
                      \kern5pt 
                      \vrule width.6pt  height5pt 
                      \kern5pt 
                      \vrule width.6pt  height5pt 
                      \kern5pt 
                      \vrule width.6pt  height5pt 
                      \kern5pt
                      \vrule width.6pt height5pt}
               \hrule height.6pt}
              }
           }}


\nref\Mumford{D. Mumford and J. Fogarty, {\it Geometrical Invariant Theory} 
  (Springer, 1982).}
\nref\LutyTaylor{M.A. Luty and W. Taylor, Phys. Rev. {\bf D53} (1996) 3399.}
\nref\SeibergI{N. Seiberg, Phys. Rev. {\bf D49} (1994) 6857.}
\nref\Pesando{I. Pesando, Mod. Phys. Lett {\bf A10} 1995, 1871.}
\nref\Giddings{S.B. Giddings and J.M. Pierre, Phys. Rev. {\bf D52} (1995) 
  6065.}
\nref\Pouliot{P. Pouliot, Phys. Lett. {\bf B359} (1995) 108.}
\nref\Gursey{F. G\"ursey, in {\it Second Workshop on Current Problems in High 
  Energy Particle Theory}, edited by G. Domokos and S. K\"oresi-Domokos, 
  (Johns Hopkins Press, Baltimore, 1977), p.~3.}
\nref\Kogan{Y.I. Kogan, A.Y. Morozov, M.A. Olshanetskii and M.A. Shifman, 
  Sov. J. Nucl. Phys. {\bf 43} (1986) 1022.}
\nref\Elashvili{A.G. Elashvili, Funk. Anal. Pril. {\bf 6} (1972) 51.}
\nref\Kephart{T.W. Kephart and M.T. Vaughn, Ann. Phys. {\bf 145} (1983) 162.}
\nref\Sikivie{F. Gursey, P. Ramond and P. Sikivie, Phys. Lett. {\bf B177} 1976.}
\nref\DeRujula{A. DeRujula, H. Georgi and S.L. Glashow, in {\it Fifth
 Workshop on Grand Unification}, edited by K. Kung, H. Fried and P. Frampton
 (World Scientific, Singapore, 1984) p.~88.}
\nref\Slansky{R. Slansky, Phys. Rept. {\bf 79} (1981) 1.} 
\nref\PatiSalam{J.C. Pati and A. Salam, Phys. Rev. {\bf D10} (1974) 275.}
\nref\tHooft{G. `t Hooft, Phys. Rev. Lett. {\bf 37} (1976) 8; Phys. Rev. {\bf 
  D14} (1976) 3432.}
\nref\Preskill{J. Preskill, S. Trivedi, F. Wilczek and M. Wise, Nucl. Phys. 
  {\bf B363} (1991) 207.}
\nref\Ibanez{L. Ib\'a\~nez and G. Ross, Phys. Lett. {\bf 260B} (1991) 291; 
  Nucl. Phys. {\bf B368} (1992) 3; L. Ib\'a\~nez, Nucl. Phys. {\bf B398} 
  (1993) 301.}
\nref\Banks{T. Banks and M. Dine, Phys. Rev. {\bf D45} (1992) 1424.}
\nref\CsakiMurayama{C. Cs\`aki and H. Murayama, hep-th 9710105.}
\nref\ADS{I. Affleck, M. Dine and N. Seiberg, Nucl. Phys. {\bf B241} (1984)
  493.}
\nref\SeibergII{N. Seiberg, Nucl. Phys. {\bf B435} (1995) 129.}
\nref\Ramond{P. Ramond, Phys. Lett. {\bf B390} (1997) 179.}
\nref\DistlerKarch{J. Distler and A. Karch, Fortsch. Phys. {\bf 45} (1997) 
  517.}
\nref\Karch{A. Karch, Phys. Lett. {\bf B405} (1997) 280.}
\nref\Leigh{R.G. Leigh and M.J. Strassler, Nucl. Phys. {\bf B496} (1997) 
  132.}
\nref\BCKS{M. Berkooz, P. Cho, P. Kraus and M. Strassler, Phys. Rev. 
  {\bf D56} (1997) 7166.}


\def\LongTitle#1#2#3#4#5{\nopagenumbers\abstractfont
\hsize=\hstitle\rightline{#1}
\hsize=\hstitle\rightline{#2}
\hsize=\hstitle\rightline{#3}
\vskip 0.5in\centerline{\titlefont #4} \centerline{\titlefont #5}
\abstractfont\vskip .3in\pageno=0}
 
\LongTitle{HUTP-97/A099}{}{}
{Moduli in Exceptional SUSY Gauge Theories}
{}{}

\centerline{Peter Cho}
\centerline{Lyman Laboratory}
\centerline{Harvard University}
\centerline{Cambridge, MA  02138}

\vskip 0.3in
\centerline{\bf Abstract}
\bigskip

	The low energy structures of $\CN=1$ supersymmetric models with
$E_6$, $F_4$ and $E_7$ gauge groups and fundamental irrep matter contents
are studied herein.  We identify sets of gauge invariant composites which
label all flat directions in the confining/Higgs phases of these theories.
The impossibility of mapping several of these primary operators rules out 
previously conjectured exceptional self duals reported in the literature.

\Date{12/97}

\newsec{Introduction}

	Supersymmetric gauge theories possess a number of general
characteristics which distinguish them from their non-supersymmetric
counterparts.  One such feature is the existence of flat directions along
which the scalar potential vanishes.  Up to local and global symmetry
transformations, all points located on flat directions represent degenerate
but physically inequivalent ground states.  These different vacua can be
described in terms of gauge dependent expectation values for matter fields
within the microscopic theory.  Alternatively, the potential's flat
directions may be labeled by gauge invariant combinations of the matter
fields \refs{\Mumford,\LutyTaylor}.  In either case, a detailed analysis of
the moduli space of vacua typically enters into the study of the infrared
dynamics of any supersymmetric model.

	Before one can begin to investigate how quantum effects may deform
or even destroy the classical low energy picture of a particular theory,
one must first identify an appropriate set of moduli space coordinates.  It
is relatively straightforward to perform this task in simple models such as
$\CN=1$ SUSY QCD with $\Nf \le \Nc+1$ flavors.  In this case, ground states
are labeled by expectation values of composite meson and baryon operators
built out of quark and antiquark superfields \SeibergI.  However, the
problem of finding a suitable set of gauge invariant moduli grows harder
for theories with more complicated color groups or matter contents.  No
prescription for how to solve this problem in an arbitrary model is known.

	In this note, we identify sets of hadrons which act as moduli space
coordinates for $\nobreak{\CN=1}$ theories based upon exceptional gauge
groups containing $\Nf$ matter superfields in the fundamental irrep.  Once
these composite operator sets are established, one can straightforwardly
analyze the vacuum structure of the models' confining/Higgs phases.
Identification of gauge invariant composites also represents a necessary
first step in the search for dual descriptions of the exceptional theories'
nonabelian Coulomb phases.  As the $G_2$ model has already been thoroughly
studied \refs{\Pesando,\Giddings,\Pouliot} while the $E_8$ model does not
confine for any nonzero value of $\Nf$, we restrict our focus to the $E_6$,
$F_4$ and $E_7$ theories.  As we shall see, the confining/Higgs phases of
these theories terminate with quantum constraints.  The physics of
confinement within the exceptional gauge group models is consequently
analogous to that for SUSY QCD with $\Nf \le \Nc$ quark flavors \SeibergI.
But their group theory structures are much more interesting and rich.

	Our article is organized as follows.  We first recall a few salient
facts about exceptional group theory in section~2.  We then construct the
composite operators which label all flat directions within the $E_6$, $F_4$
and $E_7$ models' confining/Higgs phases in section~3.  Several of these
operators had not been known before, and we demonstrate in section~4 that
their existence rules out previously conjectured self duals involving
exceptional gauge groups.  Finally, we close in section~5 with some
thoughts on constructing genuine duals to exceptional SUSY gauge theories.

\newsec{A little exceptional group theory}

	The group theory underlying the exceptional Lie algebra $E_6$ may
seem much more alien than that for the more familiar classical groups.
$E_6$ is defined in terms of $3 \times 3$ matrices with complex octonion
elements whose multiplication rule is neither commutative nor associative
\refs{\Gursey,\Kogan}.  Its 27-dimensional fundamental and 78-dimensional 
adjoint representations are large and unwieldy.  The invariant tensors of $E_6$
include a symmetric $\dabc$ symbol whose indices range over the fundamental
rather than adjoint irrep.  The generic symmetry breaking pattern
\eqn\Esixpattern{E_6 \> {\buildrel 27 \over \longrightarrow} \>
                 F_4 \> {\buildrel 27 \over \longrightarrow} \> 
                 SO(8) \> {\buildrel 27 \over \longrightarrow} \>
                 SU(3) \> {\buildrel 27 \over \longrightarrow} \>
                 1}
under which the rank-6 algebra decomposes is not intuitively obvious 
\Elashvili.  Given all these group theory difficulties, it is small wonder that
$E_6$ model building is not commonplace.

	However, a surprising amount of insight into $E_6$ can be gained by
focusing upon its regular maximal subalgebras which involve only $SU(N)$
and $Sp(2N)$ factors.  After constructing its extended Dynkin diagram and
eliminating the central root, one readily finds that $E_6$ contains an
amusing $SU(3) \times SU(3) \times SU(3) \times Z_3$ subgroup under which
the fundamental decomposes as $27 \to (3,\bar{3},1)+(1,3,\bar{3})+
({\bar3},1,3)$.  A field $Q$ that transforms according to the fundamental
irrep may consequently be regarded as either a 27-dimensional vector or as
a triplet $(L,M,N)$ of $3 \times 3$ matrices \Kephart.  Recalling the
embedding of the Standard Model inside $E_6$ Grand Unified Theories in
which the first $SU(3)$ factor is identified with color while the remaining
two are regarded as left and right handed factors
\refs{\Sikivie,\DeRujula}, we name the individual fields within the $27$ so
as to indicate their color and electric charge assignments:
\eqn\chargedLMNbasis{
L = \pmatrix{ d_1 & u_1 & D_1 \cr
	      d_2 & u_2 & D_2 \cr
	      d_3 & u_3 & D_3 \cr} 
\qquad 
M = \pmatrix{ N^0 & \bar{E} & \nu \cr
	      E & \bar{N}^0 & e \cr
	      \bar{\nu} & \bar{e} & n^0 \cr} 
\qquad
N = \pmatrix{ \dbar_1 & \dbar_2 & \dbar_3 \cr
	      \ubar_1 & \ubar_2 & \ubar_3 \cr
	      \Dbar_1 & \Dbar_2 & \Dbar_3 \cr}.}
As can be seen from these matrices, the $27$ incorporates all the familiar
members of a single Standard Model family along with some extra leptons
and charge $\third$ quarks.

	$E_6$ also contains a regular $SU(2) \times SU(6)$ maximal
subalgebra under which the fundamental breaks apart as $27 \to (1,15) +
(2,\bar{6})$.  The $27$ may therefore be written in terms of the
antisymmetric $6 \times 6$ matrix
\eqn\Amatrix{A = \pmatrix{
0    & D_3        & -u_3       & -N^0       & -\bar{E}   & -\nu     \cr
-D_3 &  0         &  d_3       & -E         & -\bar{N}^0 & -e       \cr
u_3  & -d_3       &  0         & -\bar{\nu} & -\bar{e}   & -n^0     \cr
N^0  & E          &\bar{\nu}   &  0         & \Dbar_3    & -\ubar_3 \cr
\bar{E} & \bar{N}^0  & \bar{e} & -\Dbar_3   & 0          & \dbar_3  \cr
\nu  & e          & n^0        & \ubar_3    & -\dbar_3   & 0        \cr}}
along with the $2 \times 6$ matrix 
\eqn\Bmatrix{B = \pmatrix{
d_1 & u_1 & D_1 & -\dbar_2 & -\ubar_2 & -\Dbar_2 \cr
d_2 & u_2 & D_2 &  \dbar_1 &  \ubar_1 &  \Dbar_1 \cr} .}
By examining the relationships between its two maximal subalgebras,
one can figure out much about the group theory structure of $E_6$ itself.

	For example, it is interesting to inquire how the $E_6$ invariant $Q^3
\equiv \dabc Q^\a Q^\b Q^\c$ decomposes under the $SU(3)^3$ and $SU(2)
\times SU(6)$ subgroups.  In the former case, $Q^3$ must break apart into a 
linear combination of $\det L +\det M + \det N$ and $\Tr LMN$, for only
these terms are cubic in $Q$ and invariant under the cyclic $Z_3$
permutation symmetry.  Similarly, $Q^3$ must be expressible as a linear
combination of $\Pf A$ and $\e_{ij} B^i_a A^{ab} B^j_b$.  After equating
these two linear combinations and solving for their unknown coefficients,
we find \Kephart
\eqn\dabceqn{\dabc Q^\a Q^\b Q^\c = \det L + \det M + \det N - \Tr LMN
		 = \Pf A + \half \e_{ij} B^i_a A^{ab} B^j_b.}
Numerical values for $\dabc$ may readily be computed from this result.  We
tabulate the independent, nonvanishing components of this symmetric symbol
in Appendix~A.

	Maximal regular subalgebra considerations also shed light upon
$E_6$ symmetry breaking.  As \Esixpattern\ indicates, $E_6$ breaks down to
$F_4$ when a single field $Q \sim 27$ develops a vacuum expectation value.
Working with the basis in \chargedLMNbasis, we can rotate this vev into the
form $\vev{Q} = \bigl(\vev{L},\vev{M},\vev{N} \bigr) \propto
\bigl( {\bf 0}, \bf{1}_{3 \times 3}, {\bf 0} \bigr)$ which remains
invariant under only an $SU(3)_\C \times SU(3)_{\L+\R}$ subgroup of
$SU(3)_\C \times SU(3)_\L \times SU(3)_\R$.  The nonvanishing value for
$\vev{M}$ also implies that the antisymmetric matrix $A$ in \Amatrix\
develops a vev proportional to $\sigma_2 \otimes 1_{3 \times 3}$.  The
$SU(2) \times SU(6)$ subgroup of $E_6$ consequently reduces to $SU(2)
\times Sp(6)$.  As a check, one can verify that $F_4$ contains both
$SU(3)\times SU(3)$ and $SU(2) \times Sp(6)$ as maximal regular subalgebras
under which its 26-dimensional irrep respectively decomposes as
$(3,\bar{3})+(\bar{3},3)+(1,8)$ and $(2,6) + (1,14)$ \Slansky.

	If a second $Q \sim 27$ field acquires a nonvanishing expectation
value proportional to $\bigl( {\bf 0}, \bf{\lambda_3}, {\bf 0} \bigr)$
where $\lambda_3$ denotes the third $SU(3)$ Gell-Mann matrix, $E_6$ breaks
to $SO(8)$.  The fundamental irrep then decomposes as $27 \to 8_\V + 8_\S
+8_\C + 3(1)$.  As we outline in Appendix~B, the low energy structure of
$\CN=1$ supersymmetric $SO(8)$ theories with $\Nf$ vectors, $\Nf$ spinors
and $\Nf$ conjugate spinors can be understood for any number of flavors.
It is consequently instructive to determine how $SO(8)$ is embedded within
$E_6$ in order to gain insight into models based upon the exceptional gauge
group.  We utilize the Dynkin basis matrices $P(E_6 \to F_4)$, $P(F_4 \to
SO(9))$ and $P(SO(9) \to SO(8))$ listed in ref.~\Slansky\ to project all 27
weights in the fundamental of $E_6$ down to $SO(8)$.  After matching the
results with the weights for the 8-dimensional vector, spinor and conjugate
spinor irreps, we identify how each element within $Q=(L,M,N) \sim 27$
transforms under the $SO(8)$ subgroup:
\eqn\soeightdecomp{
Q = \Biggl[
    \pmatrix{ s_2 & c_6 & v_2 \cr
	      s_3 & c_7 & v_3 \cr
	      s_5 & c_8 & v_5 \cr}, 
    \pmatrix{ \a_1  & v_8  & c_4  \cr
	      v_1   & \a_2 & s_1  \cr
	      c_5   & s_8  & \a_3 \cr}, 
    \pmatrix{ s_4 & s_6 & s_7  \cr
	      c_1 & c_2 & c_3   \cr
	      v_4 & v_6 & v_7   \cr} \Biggr]}
where $v_i \in 8_\V$, $s_i \in 8_\S$, $c_i \in 8_\C$ and $\a_1, \a_2, \a_3
\sim 1$.  Comparing this basis for $Q$ with the colored and electrically
charged one in \chargedLMNbasis, we observe that the down quarks and
electrons together with their antiparticles form complete $SO(8)$ irreps as
do the up quarks and neutrinos.  Clearly, the $SO(8)$ subgroup contains and
generalizes Pati-Salam $SU(4)$ which treats leptons as fourth colored
quarks \PatiSalam.

	As we shall see in the following sections, SUSY gauge theories
based upon $E_6$, $F_4$ and $SO(8)$ are closely linked.  We will therefore
find it useful to keep in mind the relations summarized in Table~1 between
these exceptional groups and their maximal subalgebras.

\midinsert
\parasize=1in
\def\tstrut{\vrule height 4ex depth 1.5ex width 0pt}
\begintable
$E_6$ & $\buildrel \vev{27} \over \longrightarrow$ & $F_4$ & 
$\buildrel \vev{27} \over \longrightarrow$ & $SO(8)$ \nr
$SU(3)_\C \times SU(3)_\L \times SU(3)_\R$ & $\longrightarrow$ & 
$SU(3)_\C \times SU(3)_{\L+\R}$ & $\longrightarrow$ & 
$SU(3)_\C \times U(1)_{3\V} \times U(1)_{8\V}$ \nr
$SU(2) \times SU(6)$ & $\longrightarrow$ & 
$SU(2) \times Sp(6)$ & $\longrightarrow$ & 
$SU(2)^4$ \endtable
\centerline{Table 1: Symmetry breaking patterns among exceptional groups}
\centerline{and their maximal regular subalgebras}
\endinsert
%

\newsec{Confining exceptional theories}

	We begin our study of exceptional SUSY gauge theories by
considering a model with the symmetry group
\eqna\Esixmodel
$$ \eqalignno{G &= E_6 \times \bigl[ SU(\Nf) \times U(1)_\R \times Z_{6\Nf}
\bigr]_{\rm global} & \Esixmodel a } $$
and matter content
$$ \eqalignno{
Q & \sim \bigl( 27 ; \sp \fund \sp ; R=1-4/\Nf , 1 \bigr). & \Esixmodel b}$$
Several points about this model should be noted.  Firstly, it is
reminiscent of supersymmetric quantum chromodynamics inasmuch as it
contains only fundamental irrep matter.  But since we do not choose to
incorporate antiquarks transforming as $\bar{27}$ along with the quarks in
\Esixmodel{b}, this $E_6$ model is chiral.  Secondly, the classical field
theory remains invariant under phase rotations which count quark number.
In the quantum theory, only a discrete $Z_{6\Nf}$ subgroup of $U(1)_\Q$
survives as a nonanomalous symmetry \refs{\tHooft{--}\CsakiMurayama}.
Finally, the $E_6$ model is asymptotically free so long as its one-loop
Wilsonian beta function coefficient
\foot{We adopt the $E_6$ index values $K(27) = 6$ and $K(78)=24$.  For
later use, we also record the indices $K(26)=6$ $[K(56)=12]$ and $K(52)=18$
$[K(133)=36]$ of the fundamental and adjoint irreps of $F_4$ $[E_7]$.}
\eqn\betaWilson{b_0 = \half \bigl[ 3 K({\rm Adj}) -
\sum_{\rm{\buildrel matter \over {\scriptscriptstyle reps \>\rho}}} K(\rho)
\bigr] = 3(12-\Nf)}
remains positive.  Its infrared dynamics are consequently nontrivial provided 
$\Nf < 12$.
	
	Quark field expectation values break the $E_6$ gauge symmetry at
generic points in moduli space according to the pattern displayed in
\Esixpattern.  We can use this symmetry breaking information to count the gauge
invariant operators which act as coordinates on the moduli space of
degenerate vacua for small values of $\Nf$.  In Table~2, we list the
initial parton matter degrees of freedom, the unbroken color subgroup and
the number of matter fields eaten by the superHiggs mechanism as a function
of $\Nf$.  The remaining uneaten parton fields correspond to independent
hadrons in the low energy effective theory which represent either Goldstone
bosons resulting from global $SU(\Nf)$ symmetry breaking or massless moduli
labeling D-flat directions of the $E_6$ scalar potential.  The counting
results displayed in the last column of Table~2 essentially fix the forms
of the composites within the $\Nf \le 4$ $E_6$ models:
\eqn\Esixhadrons{\eqalign{
B &= Q^3 \sim \bigl(1; \sp \symthree \sp ; 3R, 3\bigr) \cr
C &= Q^6 \sim \bigl(1; \sp \twotwotwo \sp ; 6R, 6\bigr) \cr
D &= Q^{12} \sim \bigl(1; \sp \threethreethreethree \sp ; 12R, 12 \bigr). \cr}}

\midinsert
\parasize=1in
\def\tstrut{\vrule height 4ex depth 1.5ex width 0pt}
\begintable
$\quad \Nf \quad $ \| Parton DOF \| Unbroken Subgroup \| Eaten DOF
\| Hadrons \crthick
1 \| 27 \| $F_4$   \| $78-52=26$ \| 1 \nr
2 \| 54 \| $SO(8)$ \| $78-28=50$ \| 4 \nr
3 \| 81 \| $SU(3)$ \| $78-8=70$  \| 11 \nr
4 \| 108 \|  1 \| 78 \| 30 \endtable
\centerline{Table 2:  Number of independent massless hadrons in the $E_6$ 
model}
\endinsert

	This proposed hadron set satisfies several consistency checks.  We
first recall that $E_6$ irreps belong to one of three different
``triality'' equivalence classes \Slansky.  The 27-dimensional fundamental
irrep has triality 1, whereas the singlet has triality 0.  Since all gauge
invariant composites are $E_6$ singlets, they must contain a multiple of
three quark constituents.  The hadrons in \Esixhadrons\ clearly satisfy
this necessary condition.

	We next tally the number of composite degrees of freedom as a
function of $\Nf$ in Table~3 and compare with the results of Table~2.  For
$\Nf \le 3$, $B$ and $C$ precisely account for the required number of
massless fields.  On the other hand, the hadron count exceeds the needed
number of composites by one when $\Nf=4$.  A single relation must exist
among $B$, $C$ and $D$ in this case.  The quantum constraint is restricted
by dimensional analysis, $SU(4)$ invariance and discrete symmetry
considerations.  It schematically appears in superpotential form as
\eqn\Wconstraint{\eqalign{
W_{\Nf=4} &= X \bigl[ D^2 + C^4 + C^3 B^2 + C^2 B^4 + C B^6 + B^8 
- \Lambda_4^{24} \bigr] \cr}}
where $X \sim (1; 1; 2,0)$ represents a Lagrange multiplier field.  The
undetermined numerical coefficients multiplying each term in this
expression can in principle be determined by Higgsing $E_6$ down to
$SO(8)$.  Their values are then fixed by the requirement that the exact
quantum constraint in (B.4) be recovered after the $E_6$ hadrons in
\Esixhadrons\ are decomposed in terms of the $SO(8)$ composite operators
listed in (B.3).

\midinsert
\parasize=1in
\def\tstrut{\vrule height 4ex depth 1.5ex width 0pt}
\begintable
$\quad \Nf \quad $ \| Hadrons \| $ \sp B \sp$ \| $\sp C \sp$ \| $\sp D \sp$
\| constraint \crthick
1 \| 1  \| 1  \|    \|   \|     \cr
2 \| 4  \| 4  \|    \|   \|     \cr
3 \| 11 \| 10 \| 1  \|   \|     \cr
4 \| 30 \| 20 \| 10 \| 1 \| -1  \endtable
\centerline{Table 3:  Hadron degree of freedom count}
\endinsert

	Although we have not attempted to implement this tedious procedure
to deduce the precise form of $W_{\Nf=4}$, we at least know that the
coefficient of its first term must not vanish if the first term in (B.4) is
to be recovered along the $SO(8)$ flat direction.  As a result, the point
$B=C=0$, $D \propto \Lambda_4^{12}$ lies within the deformed quantum moduli
space.  Since an $SU(4) \times U(1)_\R \times Z_{12}$ subgroup of the full
global symmetry in \Esixmodel{a}\ remains unbroken at this point, it is
instructive to compare `t~Hooft anomalies calculated in the microscopic
$E_6$ theory and the low energy sigma model.  We could compute the hadronic
global anomalies in terms of the independent fluctuations about vevs which
satisfy the constraint in \Wconstraint.  But as the quantum numbers of $X$
are precisely opposite to those of the fluctuation which is removed by the
constraint, it is easier to instead retain all components of $B$, $C$ and
$D$ and include anomaly contributions from the Lagrange multiplier field
$X$ as well.  We then find that the parton and hadron level $SU(4)^3$,
$SU(4)^2 U(1)_\R$, and $U(1)^{1,3}_\R$ global anomalies precisely match.
Moreover, the $SU(4)^2 Z_{12}$, $Z_{12}^{1,3}$, $U(1)^2_\R Z_{12}$ and
$U(1)_\R Z^2_{12}$ anomalies match modulo $12$ \CsakiMurayama.  This
anomaly agreement indicates the hadrons in \Esixhadrons\ form a complete
set of moduli which label flat directions in the $\Nf=4$ $E_6$ theory.  On
the other hand, the parton and hadron level global anomalies do not match
for $\Nf > 4$, and the disagreement cannot be eliminated by including
additional color-singlet fields into the low energy spectrum without
disrupting the $\Nf=4$ results.  So we conclude the $E_6$ model's confining
phase terminates at this stage.
\foot{As we discuss in Appendix~B, the $SO(8)$ theory which results from
Higgsing the $\Nf > 4$ $E_6$ model exists in either a nonabelian Coulomb or
free electric phase.  Since the $\Nf > 2$ $SO(8)$ theory does not confine
at the origin of moduli space, neither does its $\Nf > 4$ $E_6$ progenitor.}

	Given the hadron set in \Esixhadrons, we can investigate the low
energy structure of the $E_6$ models with $\Nf < 4$ quark flavors.
Consistency with known results along the $SO(8)$ flat direction requires
that dynamical superpotentials be generated in these theories.  Since the
$E_6$ model in \Esixmodel{}\ is chiral, we cannot simply add a mass term to
$W_{\Nf=4}$ and integrate out heavy flavors.  However, the nonperturbative
superpotentials' forms are fixed up to numerical factors by dimensional
analysis, holomorphy and symmetry considerations:
\eqn\Esixsuperpotentials{\eqalign{
W_{\Nf=1} &= \Bigl[ {\Lambda_1^{33} \over B^2} \Bigr]^{1/9} \cr
W_{\Nf=2} &= \Bigl[ {\Lambda_2^{30} \over B^4} \Bigr]^{1/6} \cr
W_{\Nf=3} &= \Bigl[ {\Lambda_3^{27} \over C^3 + C B^4 +B^6} \Bigr]^{1/3}. \cr}}
As in SUSY QCD with $\Nf < \Nc$ flavors \ADS, these quantum terms
destabilize the $\Nf < 4$ $E_6$ models' ground states.  

	Once we know the low energy $E_6$ spectrum, we can readily 
deduce the confining phase particle content of the $F_4$ model which
results from Higgsing \Esixmodel{}:
\eqn\EsixtoFfour{\qquad
\eqalign{G &= E_6  \times \bigl[ SU(\Nf) \times U(1)_\R \times Z_{6\Nf} \bigr] 
\cr
Q & \sim  (27; \sp \fund \sp ; R=1-4/\Nf, 1) \cr
& \cr
& \qquad\qquad\qquad \downarrow \sp \mu < \Lambda_{E_6}  \cr
& \qquad\quad\qquad B = \symthree \cr
& \cr
& \qquad\quad\qquad C = \twotwotwo \cr
& \cr
& \qquad\quad\qquad D = \threethreethreethree \cr}
\qquad
\eqalign{H = F_4 & \times \bigl[ SU(\Nf-1) \times U(1)_{\R'} \times 
 Z_{6(\Nf-1)} \bigr]  \cr
Q' \sim & \bigl(26; \sp \fund \sp ; R'=1-3/(\Nf-1), 1 \bigr) \cr
\Phi' \sim & \bigl(1; \sp \fund \sp ; R'=1-3/(\Nf-1), 1 \bigr) \cr
\downarrow & \sp \mu  < \Lambda_{F_4} \cr
\qquad\symthree \sp +  & \sp \sym \sp + \sp \fund \sp + \sp 1 \cr
& \cr
\qquad\twotwotwo \sp + & \sp \twotwoone \sp + \sp  \twotwo \cr
& \cr
\qquad \threethreethreethree \sp + & \sp \threethreethreetwo \sp + \sp
\threethreethreeone \sp + \sp \threethreethree. \cr}}

\bigskip\bigskip
 
\vskip -3.75 truein
$$ \eqalign{
{ \buildrel \vev{Q_{\Nf}} \over \longrightarrow} &  \cr
& \cr
& \cr
& \cr
\longrightarrow  & \cr
& \cr
\longrightarrow & \cr
& \cr
\longrightarrow & \cr
} $$

\bigskip\bigskip\noindent
It is important to note that this diagram is commutative.  The same $F_4$
hadronic spectrum is found whether one breaks $E_6$ at high energies and
then allows the unbroken $F_4$ gauge group to confine at $\mu <
\Lambda_{F_4}$ or if one instead performs a flavor decomposition of the
$E_6$ hadrons at $\mu < \Lambda_{E_6}$.
\foot{The $E_6$ and $F_4$ models' scales are related by the matching 
condition $\Lambda_{E_6}^{12-\Nf} = \vev{Q^\Nf}^3 \Lambda_{F_4}^{9-\Nf}
\propto \vev{B^{\Nf\Nf\Nf}} \Lambda_{F_4}^{9-\Nf}$.}
The two components of the $B$ baryon which transform according to the
fundamental and singlet irreps of $SU(\Nf-1)$ are respectively identified
with the color-singlet $\Phi'$ field and quark vev $\vev{Q^\Nf}$ in the
microscopic $F_4$ theory.  All the remaining hadrons on the LHS of
\EsixtoFfour\ represent composites of the 26-dimensional $Q'$ quarks.  

	As a check, one can verify that global anomaly matching demonstrates
the $\Nf-1=3$ $F_4$ spectrum is saturated by the gauge invariant moduli
\eqn\Ffourhadrons{\eqalign{
M'=\Qp^2 &\sim (1; \sp \sym \sp ; 2R', 2)  \cr
B'=\Qp^3 & \sim (1; \sp \symthree \sp ; 3R', 3) \cr
N'=\Qp^4 & \sim (1; \sp \twotwo \sp ; 4R', 4) \cr}
\qquad\qquad
\eqalign{
O'=\Qp^5 & \sim (1; \sp \twotwoone \sp ; 5R', 5) \cr
C'=\Qp^6 & \sim (1; \sp \twotwotwo \sp ; 6R', 6)  \cr
P'=\Qp^9 & \sim (1; \sp \threethreethree \sp ; 9R', 9) \cr}}
along with a field $X' \sim (1; 1; 2, 0)$.  This last object acts as a
Lagrange multiplier whose equation of motion yields the quantum constraint
\eqn\Ffourconstraint{\eqalign{
& P'^2 + O'^2 N'^2 + O'^2 M'^4 + O'^2 N' M'^2 + O' M'^5 B'\cr
& \qquad + O' M'^3 N' B' + O' M' N'^2 B' + N'^3 C' + N'^3 B'^2 + N'^2 M'^2 C
\cr
& \qquad + N'^2 M'^2 B'^2 + N' M'^4 C' + N' M'^4 B'^2 
+ M'^6 C' + M'^6 B'^2 = {\Lambda'_3}^{18} \cr}}
that comes from Higgsing the $E_6$ relation in \Wconstraint.  Since $F_4$
has only real representations, there is no group theory obstruction to
integrating out matter fields from this constraint and deriving the
dynamical superpotentials in the $F_4$ models with one and two quark
flavors.

	We conclude our survey of exceptional SUSY theories by briefly 
sketching the outlines of an $E_7$ model with nonanomalous symmetry group 
\eqna\Esevenmodel
$$ \eqalignno{G &= E_7 \times \bigl[ SU(\Nf) \times U(1)_\R \times Z_{12\Nf}
\bigr]_{\rm global}, & \Esevenmodel a} $$
matter content
$$ \eqalignno{Q  & \sim \bigl( 56; \sp \fund \sp ; R=1-3/\Nf, 1 \bigr) &
\Esevenmodel b } $$
and Wilsonian beta function coefficient $b_0 = 6
(9-\Nf)$.  At generic points in moduli space, the gauge group breaks
according to the pattern \Elashvili\
\eqn\Esevenpattern{E_7 \> {\buildrel 56 \over \longrightarrow} \>
                   E_6 \> {\buildrel 56 \over \longrightarrow} \> 
                   SO(8) \> {\buildrel 56 \over \longrightarrow} \>
                   1,}
and the pseudoreal fundamental irrep of $E_7$ decomposes as $56 \to 27 +
\bar{27} + 2(1) \to 2 (8_\V + 8_\S + 8_\C) + 8(1)$.  Using this group theory 
information to count massless degrees of freedom in exactly the same fashion 
as for the $E_6$ and $F_4$ theories, we find that the $E_7$ model's 
confining/Higgs phase spectrum looks like
\eqn\Esevenhadrons{\eqalign{
M &= Q^2 \sim ( 1; \sp \anti \sp ; 2R, 2) \cr
B &= Q^4 \sim (1; \sp \symfour \sp ; 4R, 4) \cr
C &= Q^6 \sim (1; \sp \threethree \sp ; 6R, 6) \cr
D &= Q^8 \sim (1; \sp \fourtwotwo \sp ; 8R, 8) \cr
E &= Q^{12} \sim (1; \sp \fourfourfour \sp ; 12R, 12) \cr
F &= Q^{18} \sim (1; \sp \sixsixsix \sp ; 18R, 18). \cr}}
These composites are restricted by a single quantum constraint when
$\Nf=3$.  After Lagrange multiplier contributions are taken into account,
it is again straightforward to verify that the $SU(\Nf)^3$, $SU(\Nf)^2
U(1)_\R$, $SU(\Nf)^2 Z_{12\Nf}$, $U(1)_\R^{1,3}$, $Z_{12\Nf}^{1,3}$,
$U(1)^2_\R Z_{12\Nf}$ and $U(1)_\R Z_{12\Nf}^2$ `t~Hooft anomalies
calculated at the partonic and hadronic levels match for $\Nf=3$ but
disagree for $\Nf=4$.  The $E_7$ model consequently ceases to confine at
this juncture.

\newsec{Exceptional self duals}

	Much of the progress made during the past few years in
understanding nonperturbative aspects of $\CN=1$ supersymmetric gauge
theories has stemmed from Seiberg's key insight that vacuum structures of
strongly interacting models can sometimes be described in terms of weakly
coupled duals \SeibergII.  Although a number of such strong-weak pairs have
been discovered, no systematic method for determining dual theories has so
far been developed.  Even finding new examples of duality remains a highly
nontrivial problem.  In particular, constructing general duals to models
involving exceptional gauge groups other than $G_2$ represents an
outstanding challenge.

	Within the past year, a few investigators have claimed to find
examples of $E_6$, $F_4$ and $E_7$ self duals involving special numbers of
fundamental irrep matter fields \refs{\Ramond, \DistlerKarch,
\Karch}.  The symmetry groups for both members of the alleged dual pairs
are identical, and their matter contents are nearly the same.  As a result,
`t~Hooft anomalies in the electric and magnetic theories' continuous global
symmetry groups match in a rather trivial fashion.  Moreover, a few gauge
invariant composites within the electric theory appear to map onto magnetic
counterparts.  Based upon this circumstantial evidence, these self dual
models have been conjectured to represent the first examples of $E_6$,
$F_4$ and $E_7$ duality.

	Cs\`aki and Murayama have recently called into question the
exceptional self duals' validity \CsakiMurayama.  They noticed that certain
discrete anomalies in the electric and magnetic theories do not agree.  As
these authors pointed out, this defect might conceivably be remedied
through nonperturbative generation of accidental symmetries.  The basic
veracity of the exceptional duals has therefore remained in
doubt.  Fortunately or unfortunately, we can now use our knowledge of the
low energy $E_6$, $F_4$ and $E_7$ spectrum to rule out these self duality
proposals.

	We first examine the $\Nf=6$ $E_6$ theory of Ramond \Ramond.  The 
basic structure of his dual pair is outlined below:
\eqn\Esixselfdual{
\eqalign{
G &= E_6 \times \bigl[ SU(6) \times U(1)_\R \bigr] \cr
& \qquad Q^i  \sim (27; \sp \fund \sp; \third) \cr
& \cr
& \qquad \Wtree=0 \cr}
\qquad
\eqalign{
\Longleftrightarrow & \cr
& \cr
& \cr
& \cr}
\qquad
\eqalign{
\Gdual =& E_6 \times \bigl[ SU(6) \times U(1)_\R \bigr]  \cr
& q_i \sim (27; \sp \antifund \sp ; \third) \cr
& b^{ijk} \sim (1; \sp \symthree\sp; 1) \cr
& \Wdualtree = b^{ijk} q_i q_j q_k . \cr}}
Because the electric and magnetic quarks share the same $R$-charge
assignments, $R$ symmetry matching dictates that every electric composite
$Q^n$ which is not introduced into the dual as an elementary field must be
identified with either some $q^n$ magnetic composite or else a hadron
involving dual field strength tensors.  Since $Q$ and $q$ transform
according to conjugate representations of the nonabelian flavor group, the
$Q^n \Leftrightarrow q^n$ identification can be consistent only if the
composites transform according to real irreps of $SU(6)$.  As Ramond
observed, this condition is satisfied for the $C=Q^6$ baryon in
\Esixhadrons:
\eqn\Cmatching{C = Q^6 \sim \Bigl( 1; \sp \twotwotwo \sp; 2 \Bigr) 
\qquad\Longleftrightarrow\qquad 
c=q^6 \sim \Bigl(1; \sp \twotwotwobar \sp ; 2 \Bigr).}
However, it is impossible to similarly map the $D=Q^{12}$ baryon onto
$q^{12}$.  Instead, $D$ can only be paired with a hadron containing 6
dual quarks and 2 dual field strength tensors:
\eqna\Dmatching
$$ \eqalignno{D &= Q^{12} \sim \Bigl( 1; \sp \threethreethreethree \sp; 
  4 \Bigr) 
\qquad\Longleftrightarrow\qquad 
d = q^6 \Wdual^2 \sim \bigl(1; \sp \bar{\threethree} \sp ; 4 \bigr). &
\Dmatching a} $$
Although it is not at all obvious that one can in fact construct such a 
$d$ chiral operator which is primary,  we will assume this mapping is 
possible.  We are then similarly forced to make the converse identification
$$ \eqalignno{D' &= Q^6 W^2 \sim \bigl( 1 ; \threethree; 4 \bigr) 
\qquad\Longleftrightarrow\qquad 
d' = q^{12} \sim \Bigl(1; \sp \bar{\threethreethreethree} \sp ; 4 \Bigr). &
\Dmatching b} $$

	Up to this point, Ramond's dual pair appears to pass all anomaly
matching and operator mapping tests.  But the $E_6$ self dual must also
withstand careful scrutiny along the $F_4$ and $SO(8)$ flat directions.  In
the former case, the electric and magnetic theories in \Esixselfdual\ are
deformed as follows:
\eqn\Ffourdual{
\eqalign{
H &= F_4 \times \bigl[ SU(5) \times U(1)_\R \bigr] \cr
& \qquad Q^i \sim (26; \sp \fund \sp; {2 \over 5}) \cr
& \qquad \Phi^i \sim (1; \sp \fund \sp ; {2 \over 5}) \cr
& \cr
& \cr
& \cr
& \cr
& \qquad \Wtree=0 \cr}
\qquad
\eqalign{
\Longleftrightarrow & \cr
& \cr
& \cr
& \cr
& \cr
& \cr
& \cr
& \cr}
\qquad
\eqalign{
\Hdual =& E_6 \times \bigl[ SU(5) \times U(1)_\R \bigr]  \cr
& q_i \sim (27; \sp \antifund \sp; {4 \over 15} ) \cr
& q' \sim (27; 1 ; {2 \over 3} ) \cr
& \phi^i \sim (1; \sp \fund \sp; {2 \over 5} ) \cr
& b^{ijk} \sim (1; \sp \symthree \sp; {6 \over 5}) \cr
& m^{ij} \sim ( 1; \sp \sym \sp; {4 \over 5} ) \cr
\Wdualtree = & b^{ijk} q_i q_j q_k + m^{ij} q_i q_j q' + \phi^i q_i q' q'
	+ q' q' q'. \cr}}
The dual $E_6$ baryons which couple to $b$, $m$ and $\phi$ inside the
magnetic superpotential represent redundant operators.  If one chooses to
add a mass term in order to eliminate the $\Phi \sim \phi$ singlet, the
$q_i q' q'$ composite grows heavy and should be integrated out from the
dual.  Along the $SO(8)$ flat direction, the electric theory in
\Esixselfdual\ breaks apart as
$$ I = SO(8) \times \bigl[ SU(4)_\V \times SU(4)_\S \times SU(4)_\C  \times 
  U(1)_\B \times U(1)_\Y \times U(1)_\R \bigr] $$
\eqn\SOeightelectrictheory{\eqalign{
V & \sim (8_\V; \sp \fund \sp,1,1; 0,-2,\half ) \cr
S & \sim (8_\S; 1,\sp \fund \sp ,1; 1,1,\half ) \cr
C & \sim (8_\C; 1,1,\sp \fund \sp; -1,1,\half ) \cr}
\qquad\qquad
\eqalign{
\Phi & \sim (1;  \sp \fund \sp ,1,1; 0,4,\half )\cr
\Phi' & \sim (1; 1,\sp \fund \sp ,1; 2,-2,\half )\cr
\Phi'' & \sim (1;1,1,\sp \fund \sp ; -2,-2,\half )\cr}}
$$ \Wtree = 0, $$
while the magnetic theory reduces to 
\vfill\eject

$$ \Idual = E_6 \times \bigl[ SU(4)_{\V+\S+\C} \times (Z_3)_\B \times U(1)_\R 
  \bigr]  $$
\eqn\SOeightmagnetictheory{\eqalign{
q_i & \sim (27; \sp \antifund \sp ; 0, {1 \over 6} ) \cr
q' & \sim (27; 1 ; 1, {2 \over 3} ) \cr
q'' & \sim (27; 1 ; -1, {2 \over 3} ) \cr}
\qquad\qquad
\eqalign{
\phi^i & \sim (1; \sp \fund \sp ; 0, \half ) \cr
{\phi'}^i & \sim (1; \sp \fund \sp ; 2, \half ) \cr
{\phi''}^i & \sim (1; \sp \fund \sp ; -2, \half ) \cr}
\qquad\qquad
\eqalign{
b^{ijk} & \sim (1; \sp \symthree\sp; 0, {3 \over 2}) \cr
{m'}^{ij} & \sim ( 1; \sp\sym\sp; 2, 1 ) \cr
{m''}^{ij} & \sim ( 1; \sp\sym\sp; -2, 1 ) \cr} } 
$$ \eqalignno{
\Wdualtree &= b^{ijk} q_i q_j q_k + {m'}^{ij} q_i q_j q' 
+ {m''}^{ij} q_i q_j q''
+ \phi^i q_i q' q'' + {\phi'}^i q_i q'' q'' + {\phi''}^i q_i q' q' \cr
& \qquad + q' q' q' + q'' q'' q''. \cr} $$
Only the diagonal subgroup of the $SO(8)$ model's nonabelian flavor
symmetry and a discrete $Z_3$ subgroup of its abelian baryon number remain
intact within the microscopic $E_6$ theory.  In principle, the full global
symmetry group should be restored at the infrared fixed point
\refs{\DistlerKarch,\Leigh}.

	Intractable problems with these candidate $E_6$ duals become
apparent when one tries to trace the mappings of the $D=Q^{12}$ and
$d'=q^{12}$ baryons.  Global quantum number matching dictates the 
generalizations
\eqn\Ffouropmatching{\eqalign{
Q^{12-n} & \qquad\Longleftrightarrow\qquad q^{3+n} q'^{3-n} \Wdual^2 \cr
Q^{3+n} W^2 & \qquad\Longleftrightarrow\qquad q^{12-n} {q'}^n \cr}}
of the operator identifications \Dmatching{a,b}\ in the dual pair
\Ffourdual.  These relations are not obviously incorrect.  However, the
only electric partner allowed by color, flavor, baryon number and R-charge
conservation to the $d'=q^{12}$ baryon in the magnetic $E_6$ theory of
\SOeightmagnetictheory\ is the $SO(8)$ glueball $W^2$.  This relation looks
implausible given that the $E_6$ glueball $\Wdual^2$ also maps to $W^2$.
Global quantum number matching similarly requires the pairing
\eqn\killermatch{D''=S W^2 \sim \bigl( 8_\S; 1,\sp \fund \sp ,1; 1, {5 \over 2}
\bigr) \qquad\Longleftrightarrow\qquad 
d'' = q^{11} q' \sim \Bigl(1; \sp \bar{\threethreethreetwo} \sp ; 1, 
{5 \over 2} \Bigr),}
for no other combination of $SO(8)$ vectors, spinors, conjugate spinors and
field strength tensors has the same global charges as $d''$.  But as $S
W^2$ is not gauge invariant, this last mapping is clearly impossible.

	Mapping other composites beside the $q^{12}$ type baryons in the
candidate $SO(8) \dash E_6$ dual pair is also problematic.  For instance,
the $SO(8)$ invariant $SVC$ has the same $R$ charge and number of degrees
of freedom as the $E_6$ hadrons $b$, $q^5 q'$, $q^5 q''$ and $q q' q''$
\DistlerKarch.  However, this mapping is faulty from a $(Z_3)_\B$ standpoint. 
Since all nonredundant, primary elements of the $E_6$ chiral ring cannot be 
consistently identified along the $SO(8)$ flat direction, Ramond's 
self dual pair must be rejected.

	Similar difficulties plague the $\Nf=4$ $E_7$ dual pair of 
Distler and Karch \DistlerKarch:
\eqn\Esevenselfdual{
\eqalign{
G &= E_7 \times \bigl[ SU(4) \times U(1)_\R \bigr] \cr
& \qquad Q^i  \sim (56; \sp \fund \sp ; \quarter) \cr
& \cr
& \qquad \Wtree=0 \cr}
\qquad
\eqalign{
\Longleftrightarrow & \cr
& \cr
& \cr
& \cr}
\qquad
\eqalign{
\Gdual =& E_7 \times \bigl[ SU(4) \times U(1)_\R \bigr] \cr
& q_i \sim (56; \sp \antifund \sp ; \quarter) \cr
& b^{ijkl} \sim (1; \sp\symfour\sp; 1) \cr
& \Wdualtree = b^{ijkl} q_i q_j q_k q_l. \cr}}
Among the $E_7$ hadrons listed in \Esevenhadrons, $M=Q^2 \sim \anti \sp$ ,
$C=Q^6 \sim \threethree$ and $D=Q^8 \sim \fourtwotwo$ transform according
to real representations of the $SU(4)$ flavor group.  They may consequently
be identified with $m=q^2$, $c=q^6$ and $d=q^8$ in the magnetic theory.  On
the other hand, $E=Q^{12} \sim \fourfourfour$ and $F=Q^{18} \sim
\sixsixsix$ transform according to complex $SU(4)$ irreps.  $E$ has the
same quantum numbers as $e=q^4 \Wdual^2$.  But this operator identification
suffers from the same sorts of problems along the $SO(8)$ flat direction as
we have seen for the $Q^{12}$ baryon in Ramond's $E_6$ dual pair.
Furthermore, the only possible dual counterpart to $F$ allowed by $SU(4)
\times U(1)_\R$ is $f=q^6 \Wdual^3$.  However, the electric superfield is 
bosonic whereas its magnetic partner is fermionic.  As there is no
consistent way to map the $F$ baryon within the dual $E_7$ theory, the
duality conjecture in \Esevenselfdual\ is fatally flawed.

	All similar exceptional self dual pairs which have been reported in
the literature can be ruled out in an analogous fashion.  These invalid 
self dual examples serve as useful reminders that `t~Hooft anomaly matching 
represents a necessary but insufficient condition for establishing duality.

\newsec{Conclusion}

	With the demise of the self dual proposals, virtually nothing is
known about the existence of duals to $\CN=1$ supersymmetric theories based
upon exceptional groups other than $G_2$.  No viable magnetic counterparts
to $E_6$, $F_4$ or $E_7$ models with vanishing tree level superpotentials
have so far been uncovered.  However, careful study of the $SO(8)$ flat
directions that exist in all these theories may yield some valuable clues.
In particular, looking for generalizations of the magnetic $SU(3\Nf-5)
\times Sp(2\Nf-2)$ counterpart to the $SO(8)$ model with $\Nf$ vectors,
$\Nf$ spinors and $\Nf$ conjugate spinors outlined in Appendix~B represents
an obvious starting point in the search for exceptional duals.  It is
important to note that only an $SU(\Nf) \times SU(2)$ subgroup of the
$SO(8)$ theory's $SU(\Nf)^3$ flavor symmetry is realized at short distances
in its magnetic partner.  This suggests that the full nonabelian flavor
groups in the $E_6$, $F_4$ and $E_7$ models which we have studied might
also not exist at all energy scales within their duals.  Instead, we
believe it is more likely that only an $SU(2)$ subgroup embedded inside
$SU(\Nf)$ such that $\Nf \to \Nf$ is realized in the ultraviolet.  If this
conjecture is incorrect, matching global anomalies and operator composites
between the microscopic exceptional gauge theories and their mystery dual
partners looks formidably difficult.

	We close by noting an intriguing possibility regarding the phase
structure of the $E_6$ model in \Esixmodel{}.  Since the nonanomalous
R-charge assignment for the matter fields is unique, the scaling dimensions
of all composite operators are fixed by the relation $D = 3R/2$.  This
simple formula implies $D(B=Q^3) = 9/10$ in the $\Nf=5$ model.  General
lore holds that a magnetic dual becomes free at long distances when the
mass dimensions of all electric composites included into the dual as
fundamental fields reduce to less than unity \SeibergII.  So if $B$ is the
only elementary composite within the dual, the $E_6$ theory with $\Nf=5$
flavors may exist in a free magnetic phase.  Of course, if any parts of the
$C=Q^6$ or $D=Q^{12}$ hadrons enter into the dual as well, then the $\Nf=5$
model cannot be free since these fields' dimensions are greater than unity.
We note that certain portions of $C$ possess exactly the right symmetry
properties to be identified with bilinear meson fields along the $SO(8)$
flat direction, and such mesons appear in the magnetic superpotential of
the $SU(3\Nf-5) \times Sp(2\Nf-2)$ theory.  So the existence or absence of
a free magnetic phase for the $E_6$ model will only be known for certain
after a genuine dual is found.

\bigskip
\centerline{{\bf Acknowledgments}}
\bigskip

        I thank Per Kraus for stimulating my interest in exceptional SUSY 
models and Howard Georgi for many enlightening group theory discussions.  This 
work was supported by the National Science Foundation under Grant 
\#PHY-9218167.

\appendix{A}{Component values of the $\pmb{E_6}$ $\pmb{\dabc}$ symbol}

	We tabulate here the independent, nonvanishing components of the
totally symmetric $E_6$ $\dabc$ symbol in the basis where the $27$
decomposes under $SU(3)^3$ as $Q=(L,M,N)$ with
\eqn\hermitianLMNbasis{
L = \pmatrix{ Q^1 & Q^2 & Q^3 \cr
	      Q^4 & Q^5 & Q^6 \cr
	      Q^7 & Q^8 & Q^9 \cr} 
\qquad 
M = \pmatrix{ Q^{10} & Q^{11} & Q^{12} \cr
	      Q^{13} & Q^{14} & Q^{15} \cr
	      Q^{16} & Q^{17} & Q^{18} \cr} 
\qquad
N = \pmatrix{ Q^{19} & Q^{20} & Q^{21} \cr
	      Q^{22} & Q^{23} & Q^{24} \cr
	      Q^{25} & Q^{26} & Q^{27} \cr}.}
After inserting these forms for $L$, $M$ and $N$ into \dabceqn, we can
simply read off the values for $\dabc$.  The nonzero components all equal
either $1$ or $-1$ as indicated in the tables below:

\midinsert
\parasize=1in
\def\tstrut{\vrule height 4ex depth 1.5ex width 0pt}
\begintable
\multispan{3}\tstrut\hfil $\dabc=1$ \hfil\cr
1,5,9  | 10,14,18  | 19,23,27 \cr
2,6,7  | 11,15,16  | 20,24,25 \cr
3,4,8  | 12,13,17  | 21,22,26 \endtable
\endinsert

\midinsert
\parasize=1in
\def\tstrut{\vrule height 4ex depth 1.5ex width 0pt}
\begintable
\multispan{9}\tstrut\hfil $\dabc=-1$ \hfil\cr
1,6,8    |  2,4,9    |  3,5,7   | 4,10,20 | 5,13,20 | 6,16,20 | 7,11,24  |
8,14,24  |  9,17,24  \cr
1,10,19  |  2,13,19  |  3,16,19 | 4,11,23 | 5,14,23 | 6,17,23 | 7,12,27  |
8,15,27  |  9,18,27  \cr
1,11,22  |  2,14,22  |  3,17,22 | 4,12,26 | 5,26,15 | 6,18,26 | 10,15,27 |
11,13,18 |  12,14,16 \cr  
1,12,25  |  2,15,25  |  3,18,25 | 7,10,21 | 8,13,21 | 9,16,21 | 19,24,26 |
20,22,27 | 21,23,25  \endtable
\endinsert
\noindent
It is amusing to note that only $45 \times 3!=270$ components of $\dabc$
out of the possible $27^3=19,683$ are nonvanishing in the basis
\hermitianLMNbasis.  

\appendix{B}{$\pmb{SO(8)}$ theory with equal numbers of vector, spinor and 
conjugate spinor fields}

	The $E_6$ model in \Esixmodel{}\ with $\Nf+2$ fundamental matter
fields reduces to an $SO(8)$ theory with $\Nf$ vectors, $\Nf$ spinors and
$\Nf$ conjugate spinors at points in moduli space where two of the $27$'s
develop nonvanishing expectation values.  The same $SO(8)$ theory can be
reached by starting from an $SO(10)$ model with $\Nf+2$ 10-dimensional
vectors and $\Nf$ 16-dimensional spinors.  Since duals to $SO(10)$ SUSY
gauge theories containing arbitrary numbers of vectors and spinors have
recently been uncovered \BCKS, the vacuum structure of the $SO(8)$
model can be studied for all values of $\Nf$.  We sketch its principle
features in this appendix.

	The microscopic $SO(8)$ model has the full symmetry group 
\eqn\soeightsymgroup{G = SO(8) \times \bigl[SU(\Nf)_\V \times SU(\Nf)_\S 
\times SU(\Nf)_\C \times U(1)_\B \times U(1)_\Y \times U(1)_\R 
\bigr]_{\rm global} }
and matter content
\eqn\soeightmatter{\eqalign{
V^i & \sim \Bigl( 8_\V; \sp \fund \sp ,1,1; 0,-2, R \Bigr) \cr
Q^\I_\S  & \sim \Bigl( 8_\S; 1,\sp \fund \sp ,1; 1, 1,  R \Bigr) \cr
Q^\Idot_\C  & \sim \Bigl( 8_\C; 1,1,\sp \fund \sp ;-1, 1,  R \Bigr) \cr}}
where $R = 1-2/\Nf$.  This theory confines everywhere throughout moduli space 
for $\Nf \le 2$.  The gauge invariant composites
\eqn\soeighthadrons{
\eqalign{
M^{(ij)}  &= V^i V^j \cr
\LS^{(\I\J)} &= Q_\S^\I C Q_\S^\J \cr
\LC^{(\Idot\Jdot)} &= Q_\C^\Idot C Q_\C^\Jdot \cr
N^{\I i \Idot}  &= Q_S^\I \Vslash^i C Q_\C^\Idot \cr}
\qquad\qquad
\eqalign{
{\OSV}^{[ij]}_{[\I\J]} &= {1 \over 2!} {Q_\S}_{[\I} \Vslash^{[i} \Vslash^{j]} 
  C {Q_\S}_{\J]} \cr
{\OCV}^{[ij]}_{[\Idot\Jdot]} &= {1 \over 2!} {Q_\C}_{[\Idot} \Vslash^{[i} 
  \Vslash^{j]} C {Q_\C}_{\Jdot]} \cr
{\OSC}^{[\I\J]}_{[\Idot\Jdot]} &= {1 \over 2!} {Q_\S}^{[\I} \Gamma^\mu C 
{Q_\C}_{[\Idot} {Q_\S}^{\J]} \Gamma_\mu C {Q_\C}_{\Jdot]} \cr
P^{[ij][\I\J]}_{[\Idot\Jdot]} &= {1 \over 2!} {Q_\S}^{[\I} \Gamma^\mu C 
 {Q_\C}_{[\Idot} {Q_\S}^{\J]} \Gamma_\mu \Vslash^{[i} \Vslash^{j]} C 
 {Q_\C}_{\Jdot]} \cr}}
act as confining phase moduli space coordinates.  In the $\Nf=2$ model,
these hadrons are not all independent.  Instead, they are related by the
single quantum constraint
\eqn\soeightconstraint{\eqalign{
& P^2 - 4 \OSC \OSV \OCV + 8 N^4
   + 2 \bigl[ M^2 \OSC^2 + \LS^2 \OCV^2 + \LC^2 \OSV^2 \bigr] \cr
& \qquad - 8 \bigl[ N^2 M \OSC + N^2 \LS \OCV + N^2 \LC \OSV \bigr] 
   + 16 N^2 M \LS \LC - 2 M^2 \LS^2 \LC^2 = \Lambda^{12}_2.  \cr}}
One can check that this exact relation reduces to the correct constraint for
$\Nf=\Nc=4$ SUSY QCD along the flat direction where vevs for the two
vectors break $SO(8)$ down to $SO(6) \simeq SU(4)$.

	For $3 \le \Nf \le 5$, the $SO(8)$ theory no longer confines at the 
origin of moduli space.  Instead, it exists in a nonabelian Coulomb phase 
which can be described in terms of a dual model with the symmetry group 
\eqn\maggroup{\Gdual = \bigl[SU(\Ncdual) \times Sp(2\Ncpdual) \bigr]_{\rm
local} \times \bigl[ SU(\Nf) \times SU(2) \times U(1)_\B \times U(1)_\Y 
\times U(1)_\R \bigr]_{\rm global}}
where $\Ncdual = 3\Nf-5$ and $\Ncpdual = \Nf-1$.  The dual has matter 
content 
\eqn\magmatter{
\eqalign{
q &\sim \bigl( \sp \fund \sp ,1 ; \sp \antifund \sp,1; 0, Y_q, R_q \bigr) \cr
\qp &\sim \bigl( \sp \fund \sp ,\sp\fund\sp ; 1,2; 0, Y_{q'}, R_{q'} \bigr) \cr
\qpp &\sim \bigl( \sp\fund\sp,1; 1,1; -2, Y_\qpp, R_\qpp \bigr) \cr
\qppp &\sim \bigl( \sp\fund\sp,1; 1,1; 2, Y_\qppp, R_\qppp \bigr) \cr
\qbar &\sim \bigl( \sp \antifund \sp ,1; 1,2\Nf-1; 0,Y_\qbar, R_\qbar \bigr) \cr
s &\sim \bigl( \sp\symbar\sp,1; 1,1; 0, Y_s, R_s \bigr) \cr} 
\qquad 
\eqalign{
t &\sim \bigl( 1,\sp\fund\sp; 1,2\Nf-2; 0,Y_t, R_t \bigr) \cr
m &\sim \bigl(1,1; \sp\sym\sp,1; 0, Y_m, R_m \bigr) \cr
\ell_\S &\sim \bigl(1,1; 1,2\Nf-1; 2, Y_{\ell_\S}, R_{\ell_\S} \bigr) \cr
\ell_\C &\sim \bigl(1,1; 1,2\Nf-1; -2, Y_{\ell_\C}, R_{\ell_\C} \bigr) \cr
n &\sim \bigl(1,1; \sp\fund\sp,2\Nf-1; 0, Y_n, R_n \bigr) \cr}}
where the magnetic fields' hypercharge and R-charge assignments are given
by
\eqn\magcharges{\eqalign{
Y_q & = {4 \Nf (\Nf-2)  \over \Ncdual} \cr
Y_{q'} = Y_\qpp = Y_\qppp &= -{2 \Nf(\Nf-1) \over \Ncdual} \cr
Y_\qbar & =   -{4 \Nf(\Nf-2) \over \Ncdual} \cr
Y_s & = {4\Nf(\Nf-1) \over \Ncdual} \cr
Y_t = Y_{\ell_\S} = Y_{\ell_\C} & = 2 \Nf   \cr
Y_m & = -4\Nf \cr
Y_n & = 0 \cr}
\qquad\qquad
\eqalign{
R_q &= - {2\Nf-7 \over \Ncdual} R   \cr
R_{q'} = R_\qpp = R_\qppp &= {\Nf+2 \over \Ncdual} R   \cr
R_\qbar &=   -{\Nf^2-12\Nf + 16 \over \Nf \Ncdual} \cr
R_s &= {4\Nf^2 - 10 \Nf + 8 \over \Nf \Ncdual} \cr
R_t = R_{\ell_\S} = R_{\ell_\C} &= 2R \cr
R_m &= 2R  \cr
R_n &= 3R. \cr}}
The magnetic theory also has a tree level superpotential which
schematically looks like
\eqn\Wmagnetic{\tilde{W} = m q s q + \qpp s \qppp  + n q \qbar 
+ \ell_\S \qpp \qbar + \ell_\C \qppp \qbar + \qp s \qp + \qp \qbar t.}
The electric $SO(8)$ and magnetic $SU(\Ncdual) \times Sp(2\Ncpdual)$ pair 
satisfy all the standard anomaly matching, operating mapping and
confinement recovery tests of duality \SeibergII.

	Finally for $\Nf \ge 6$, the $SO(8)$ model loses asymptotic freedom
and becomes a free field theory in the far infrared.  

\listrefs 
\bye